\documentclass[12pt,epsf]{article}
\usepackage{amssymb}
\usepackage{epsf}

\def\be{\begin{equation}}
\def\ee{\end{equation}}
\def\bea{\begin{eqnarray}}
\def\eea{\end{eqnarray}}

\renewcommand{\theequation}{\arabic{section}.\arabic{equation}}

\def\case#1/#2{\textstyle\frac{#1}{#2}}
\def\k0{\kappa_{0}}

\begin{document}
\begin{titlepage}
 
\vspace{.7in}
 
\begin{center}
\Large
{\bf Effects of anisotropy and spatial curvature 
on the pre-big bang scenario}
 
\vspace{.7in}
 
\normalsize
 
\large{Dominic Clancy$^{1a}$, James E. Lidsey$^{2b}$ 
\& Reza Tavakol$^{1c}$}
 
\normalsize
\vspace{.7in}
 
$^1${\em Astronomy Unit, School of Mathematical Sciences,  \\
Queen Mary \& Westfield College, Mile End Road, LONDON, E1 4NS, U.K.}

\vspace{.2in}
$^2${\em Astronomy Centre and Centre for Theoretical Physics, \\
University of Sussex, BRIGHTON, BN1 9QH, U.K.}

\end{center}
 
\vspace{.3in}
 
\baselineskip=24pt
\begin{abstract}
\noindent

A class of exact, anisotropic 
cosmological solutions to the vacuum 
Brans--Dicke theory of gravity is considered 
within the context of the pre--big 
bang scenario. Included in this class are
the Bianchi type III, V and ${\rm VI}_h$ models 
and the spatially isotropic, negatively curved 
Friedmann--Robertson--Walker universe. 
The effects of large anisotropy and 
spatial curvature are determined. 
In contrast to negatively curved Friedmann--Robertson--Walker
model, there exist regions of the parameter space 
in which the combined effects of curvature and anisotropy
prevent the occurrence of inflation.
When inflation 
is possible, the necessary and sufficient conditions 
for successful pre--big bang inflation 
are more stringent than in the isotropic models. 
The initial state for these models is established 
and corresponds in general to a gravitational plane wave. 

\end{abstract}
 
PACS NUMBERS:  04.20.Jb, 04.50.+h, 11.25.Mj, 98.80.Cq

\vspace{.3in}
$^a$Electronic address: dominic@maths.qmw.ac.uk
 
$^b$Electronic address: jel@astr.cpes.susx.ac.uk
 
$^c$Electronic address: reza@maths.qmw.ac.uk

\end{titlepage}
 
\section{Introduction}

\setcounter{equation}{0}
 
\def\theequation{\thesection.\arabic{equation}}
 
The inflationary paradigm of the very early universe 
resolves a number of problems with the hot big bang
model \cite{inflation}. In the standard,
chaotic scenario, the inflationary (accelerated) expansion
is driven by the potential energy of a self--interacting
quantum scalar field \cite{linde}. Recently, an 
alternative inflationary cosmology  
-- the {\em pre--big bang} scenario --
has been developed within the context of string theory \cite{PBB}. 
The fundamental postulate of pre--big bang cosmology is that 
the initial state of the universe should correspond to the 
string perturbative vacuum. Such a state is unstable  
to small fluctuations in the dilaton field, however, due to the 
non--minimal coupling of this field to the graviton. 
The kinetic energy of the dilaton then
drives a superinflationary expansion, 
where the Hubble radius decreases with cosmic time, and 
the universe evolves from a region of weak coupling and low 
space--time curvature
to one of strong coupling and high curvature. 
To lowest--order in the string effective action, the end state
is a singularity both in the curvature and coupling, but it is
anticipated that this will be avoided 
at the string scale  
when  higher--order terms in the action become relevant  \cite{sing}.
An important feature of this scenario, therefore, 
is that the end of inflation is determined by 
quantum gravitational effects, 
in contrast to models driven by potential energy. 

Given the potential importance of such a scenario,
it is necessary to establish its generality with 
respect to anisotropy and spatial curvature, since 
small anisotropies and curvature are bound to 
be present in the real universe. 
A number of authors have recently addressed related
questions. Veneziano and collaborators considered
the general effects of small anisotropy and inhomogeneity 
by assuming the initial
space and time
derivatives of the fields were  
vanishingly small but of the same order 
\cite{ven,ven1}. In 
this sense, the initial state of the universe 
is arbitrarily near to Minkowski space--time, but it 
is not necessarily homogeneous. It was found that 
sufficiently smooth regions inflate. 

On the other hand, 
Turner and Weinberg investigated large
spatial curvature in the isotropic Friedmann--Robertson--Walker 
(FRW) universes \cite{tw}. They concluded that curvature terms 
postpone the onset of inflation and can prevent 
sufficient inflation from occurring 
before higher--order terms become significant. This  
was interpreted as a restriction on the 
set of possible initial conditions. 
Kaloper, Linde and Bousso have also argued that initial 
conditions are severely constrained in the FRW models  
\cite{klb}. A Hamilton--Jacobi 
approach to inhomogeneous string cosmology has been 
developed \cite{Say}
and higher--order corrections in spatially curved and 
anisotropic models have also been considered \cite{mag}. Finally, a 
number of exact anisotropic and inhomogeneous 
solutions have been found \cite{ed,batakis,barrow,fen}. 

The purpose of the present paper is to investigate the combined
effects of spatial anisotropy and curvature in homogeneous,
Brans--Dicke cosmologies.  The Brans--Dicke theory includes
the dilaton--graviton sector of the string effective action as a
special case \cite{bd,string}.  We analyze the class of Bianchi type
${\rm VI}_{h}$ $(h<0)$ universes. These cosmologies are
interesting for a number of reasons.  They are amongst those 
Bianchi models with a positive
measure in the homogeneous initial data
and admit some of the most general, vacuum
solutions \cite{siklos78}. They have a positive
measure in the homogeneous initial data. 
Exact type ${\rm VI}_{h}$ solutions
with a non--trivial dilaton field have been derived previously
\cite{lp} and these solutions include the Bianchi types III and V and
negatively curved, isotropic FRW models as special cases. Direct
comparisons with the isotropic models can therefore be made by
employing these solutions and, since they are exact, the effects of
large deviations from the flat FRW model can also be
determined. The initial state for a pre--big bang scenario 
can also be found analytically. 

The paper is organised as follows. In Section 2
we present the cosmological solutions 
in the string frame, where fundamental strings trace 
geodesic surfaces with respect to the metric. We then 
determine whether large anisotropies and spatial curvature 
can prevent pre--big bang inflation. 
We derive quantitative bounds on the curvature and coupling  
for successful inflation in Section 3 and 
consider the early--time limits of the models 
in Section 4. We conclude with a 
discussion in Section 5. 

Unless otherwise stated, units are chosen such that $\hbar =c =1$.

\section{Pre--Big Bang Inflation in Bianchi Type 
${\rm VI}_{h}$ Cosmologies}

\setcounter{equation}{0}
 
\def\theequation{\thesection.\arabic{equation}}

The Bianchi cosmologies are 
spatially homogeneous models  with a line 
element given by $ds^2=-dt^2+h_{ij}(t)\omega^i \omega^j$
$(i,j=1,2,3)$, 
where $h_{ij} (t)$ is the metric on the surfaces 
of homogeneity and $\omega^i$ are one--forms \cite{ryan}. 
A three--dimensional Lie group of isometries acts 
simply--transitively on the space--like, three--dimensional 
orbits and the structure constants,  ${C^a}_{bc}$, 
of the Lie algebra of the group 
determine the isometry of the three--surfaces. In general, 
these may be expanded in terms of 
a diagonal, symmetric matrix, $m^{ab}$, such that
${C^a}_{bc} =m^{ad}\epsilon_{dbc}
+{\delta^a}_{[b}a_{c]}$, where $a_c \equiv {C^a}_{ac}$
and indices are raised with $h^{ab}$ \cite{em69}. It then 
follows from the Jacobi identity, ${C^a}_{b[c} 
{C^b}_{de]}\equiv 0$, that $m^{ab}$ and $a_b$ must be orthogonal, 
$m^{ab}a_b =0$. A specific model belongs to the Bianchi 
class A or B if $a_b=0$ or $a_b \ne 0$, respectively \cite{em69}. 
Furthermore, each equivalence class of tensors ${C^a}_{bc}$ form
a linear submanifold of the space of all 3--index tensors,
where the dimension of each submanifold (with the upper
bound of six) may be taken as a measure of that Bianchi class
within the space of initial data. The class B Bianchi type ${\rm VI}_{h}$
models (with group parameter $h<0$)
considered here are among the Bianchi models
that possess the maximum dimension 6.
When $h=-1$, this model corresponds to the Bianchi 
type III. The case $h=-1/9$ is exceptional  because $a^b$ is 
not a shear eigenvector in this case \cite{em69}.
The diagonal Bianchi type ${\rm VI}_{h}$ 
metric with ${m^a}_a=0$ is 
\be
\label{6hmetric}
ds^2 =-dt^2 + a_1^2 (t)dx^2 +a^2_2 (t)e^{-2(1+k)x} dy^2 + 
a_3^2 (t) e^{2(k-1)x} dz^2   ,
\ee
where $k \equiv 
(-h)^{-1/2}$.

The gravitational sector of the Brans--Dicke theory
of gravity is given by \cite{bd}
\be
\label{bdaction}
S=\int d^4 x\, \sqrt{-g}\, e^{-\phi} \left[ R -\omega \left( \nabla \phi
\right)^2 \right]\, ,
\ee
where $R$ is the Ricci curvature scalar of the spacetime with metric
$g_{\mu\nu}$ and signature $(-, +, +, +)$,
$g \equiv {\rm det} g_{\mu\nu}$ and  $\phi$ is the dilaton field. 
The constant parameter $\omega$ determines the coupling between 
the dilaton and graviton and  $\omega =-1$ corresponds to 
the string effective action \cite{string}. The dilaton is assumed 
to be constant on the surfaces 
of homogeneity and the type ${\rm VI}_h$ 
cosmological  
field equations derived from Eq. (\ref{bdaction})   have the form
\bea
\label{f1}
\partial_t {H}_1+3HH_1 -2  (1+k^2) a_1^{-2} + 
\partial_t ( \ln \Phi) 
\partial_t ( \ln a_1)  =0 \nonumber \\
\partial_t {H}_2 +3HH_2-2(1+k) a_1^{-2}  +\partial_t 
 (\ln \Phi ) \partial_t ( \ln a_2 ) =0 \nonumber \\
\partial_t {H}_3 +3 HH_3 -2  (1-k)a_1^{-2} +
\partial_t ( \ln \Phi ) 
\partial_t ( \ln a_3 ) =0 \nonumber \\
2H_1  = (1+k) H_2 +(1-k)H_3  \nonumber \\
H_1H_2 +H_1 H_3 +H_2H_3 - (3+k^2)a_1^{-2} +3H 
\partial_t (\ln \Phi ) 
-\frac{\omega}{2} \left[ \partial_t ( \ln \Phi ) \right]^2 = 0\nonumber \\
\label{f5}
\partial_t \left( a^3 \partial_t \Phi \right) =0   ,
\eea
where $\Phi \propto e^{-\phi}$ is the Brans-Dicke  
field, $a \equiv (a_1a_2a_3)^{1/3}$ represents an averaged 
scale factor, $H_i \equiv 
\partial_t  (\ln a_i )$, $H \equiv \left( \sum_{i} 
H_i\right) /3 $ and $\partial_t$ 
denotes differentiation with respect to cosmic time, $t$. 
The Brans--Dicke field determines the Planck length, $\Phi = l^{-2}_{\rm Pl}$, 
and, in string theory, is related to the string scale, 
$l_{\rm st}$, by $\Phi = l^{-2}_{\rm st}e^{-\phi}$.  

The field equations (\ref{f5}) may be written as \cite{lp}
\bea
\label{sep1}
g'' -4 g& =& 0 \\
\label{sep2}
( \ln y)' +y (\ln g )' &=& 2 (1+k) \\
\label{sep3}
a_2 a_3 \Phi'  &=& -B   ,
\eea
where 
\bea
g &\equiv& a_2 a_3 \Phi \\
\label{y}
y &\equiv& (\ln a_2 )'\\
\eta &\equiv& \int^t \frac{dt_1}{a_1 (t_1)},
\eea
a prime denotes differentiation with respect to
$\eta$ and $B$ is an arbitrary
constant \cite{lp}.
Eqs. (\ref{sep1})--(\ref{y}) admit the class of solutions
\cite{lp} 
\bea
\label{sol1}
a_1^2 &=& A_1^2 \left[ {\rm sinh} (-2 \eta )\right]^{1+k^2} 
\left[ {\rm tanh} (-\eta )\right]^{mk - p(1-k)} \\
a_2^2 &=& A^2_2 \left[ {\rm sinh} (-2 \eta )\right]^{1+k}
\left[ {\rm tanh} (-\eta )\right]^m \\
\label{sol3}
a_3^2 &=&A_3^2 \left[ {\rm sinh} (-2 \eta )\right]^{1-k} \left[ 
{\rm tanh} (-\eta )\right]^{-(m+2p)} \\
\label{sol4}
\Phi &=&Q^2 \left[ {\rm tanh} (-\eta ) \right]^{p}  ,
\eea
where $A_i$ and $Q$ are arbitrary constants and 
the constants $\{ m,p,k,\omega \}$ satisfy the constraint 
\be
\label{ellipse}
3+k^2 -m^2 -2p \left[ (\omega +2)p+m \right] =0  .
\ee
A linear translation on the time parameter has been 
performed without loss of generality such that the singularity 
is located at $\eta =0$. We consider solutions defined 
over negative time within the context 
of the pre--big bang scenario. 

Solution (\ref{sol1})--(\ref{sol4}) 
includes the Bianchi type III, Bianchi type V 
and negatively curved FRW models
as special cases, corresponding to $k=1$ $(h=-1)$,
$k=0$ $(h = -\infty )$ and  $(k=0, p=-m)$,
respectively. It corresponds to the Ellis--MacCallum vacuum 
${\rm VI}_{h}$ solution of 
Einstein gravity when the Brans--Dicke field is constant 
$(p=0)$ \cite{em69,m71}. In the limit $\eta 
\rightarrow 0^-$, it reduces to 
the Bianchi type I cosmology \cite{meuller}:
\be 
\label{I}
a_i \propto (-\eta)^{p_i},\qquad \Phi \propto  (-\eta )^{p}  ,
\ee
where the constants $p_i$ are defined by
\bea
\label{pconstraints}
p_1 &\equiv& \frac{1}{2} \left( 1+k^2 +mk -
p(1-k) \right) \nonumber \\
p_2 &\equiv & \frac{1}{2} (1+ k+m) \nonumber \\
p_3 &\equiv &\frac{1}{2} (1-k-2p-m)
\eea
and satisfy
\be
\label{sum}
\sum_{i=1}^3 p_i= 1+p_1 -p  .
\ee

The characteristic feature of  
pre--big bang cosmology is that the universe 
should exhibit superinflationary expansion  simultaneously in 
all three directions 
as $t \rightarrow 0^-$ $(\eta 
\rightarrow 0^-)$, i.e., $H_i>0$ and 
$ \partial_t{H}_i >0$. 
We now consider the range of parameter 
values $\{ k, m, p \}$ where this behaviour is possible. 
Due to continuity, it is sufficient 
to consider the necessary 
and sufficient conditions on the type I solution 
(\ref{I}). These are that 
$p_i < 0$ and $p_1 > -1$ and this implies that 
\bea
\label{const1}
1< p(1-k) -k(k+m) <3 \nonumber \\
\label{const2}
1+k+m <0 \nonumber \\
\label{const3}
k+m+2p >1   .
\eea
Thus, $p>0$ and the Brans--Dicke field  must be 
a monotonically decreasing function of $\eta$. 

The relevant regions in the $(m,p)$ 
plane for the type III and V models 
are illustrated in Figs. 1 and 2, respectively. 
For a given model, pre--big bang solutions are constrained 
to lie in the interior of the bold region
defined by the constraints (\ref{const3}). In addition, 
physical solutions must satisfy Eq. (\ref{ellipse}). 
In the range $-3/2 <\omega \le 0$ (which includes the string case
$\omega =-1$) and for all $k$,
Eq. (\ref{ellipse}) represents an ellipse.
The superinflationary
solutions lie along the open arcs 
bounded by the points $AC$. 
In the type V model, the region of the ellipse 
consistent with the above constraints includes the point $B$ 
representing the negatively curved
FRW solution $(p=-m=\sqrt{3})$. Although Eq. (\ref{ellipse}) 
implies that all type V solutions 
satisfy  $|p|< \sqrt{3}$, only a very narrow 
region of the ellipse, $(1+\sqrt{5})/{2} < p< \sqrt{3}$, leads to 
pre--big bang inflation. Similar considerations  
give the allowed values of $m$ 
in this model to be $ -\sqrt{5} <m<-1$.

\begin{figure}
\centerline{
\def\epsfsize#1#2{0.7#1}\epsffile{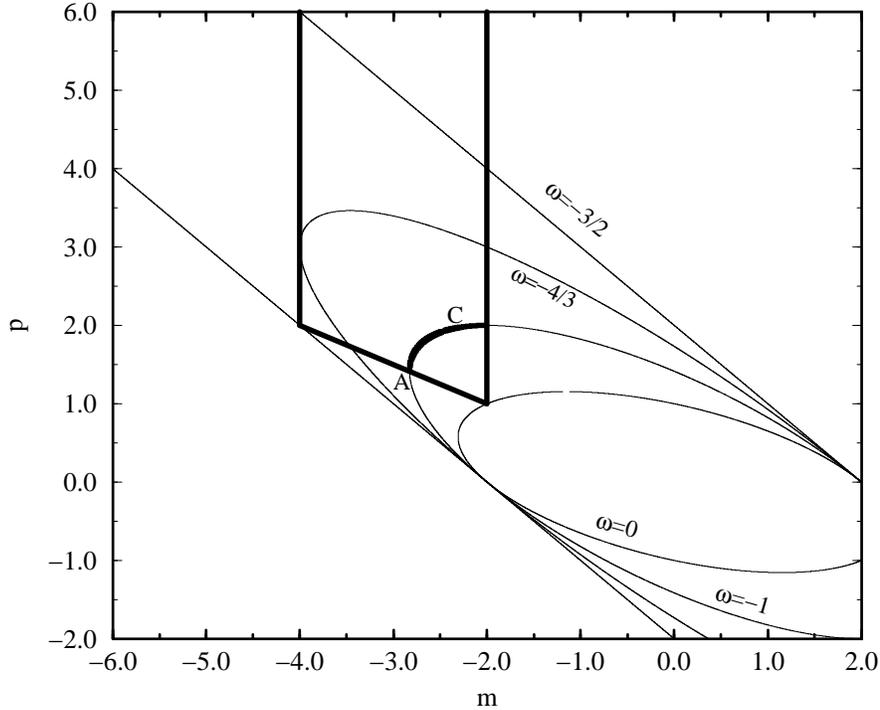}} 
\caption[]{\label{fig2}
Constraints on the parameters $m$ and $p$ permitting 
pre--big bang inflation in the Bianchi 
type III models. For $\omega =-1$, 
the bounds are
$-2\sqrt{2} <m<-2$ and $\sqrt{2}<p<2$. 
The allowed region satisfying all 
constraints is given by the arc $AC$.}
\end{figure}

\begin{figure}
\centerline{
\def\epsfsize#1#2{0.7#1}\epsffile{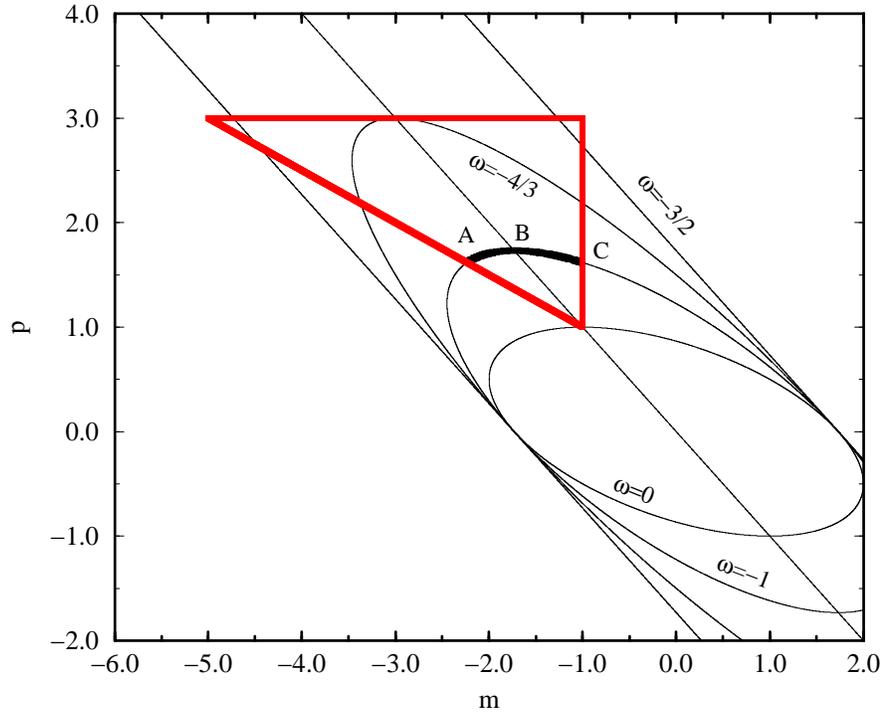}}
\caption[]{\label{fig1}
Constraints on the parameters $m$ and $p$ permitting 
pre--big bang inflation in the Bianchi 
type V models. The values of $m$ and $p$ 
must lie within the region bounded by $1< p< 3$, 
$m<-1$ and $m+2p>1$. The arc $AC$ denotes the region 
satisfying all constraints in the string model 
and includes the point $B$ 
corresponding to  the negatively curved FRW solution.}
\end{figure}

Thus, 
pre--big bang inflation is possible in these Bianchi models, 
but the combined roles of anisotropy and spatial
curvature can prevent inflation from occurring. 
This is in contrast to the 
negatively curved FRW universe, where a superinflationary 
expansion  towards a singular state is only delayed 
by the spatial curvature \cite{tw}. 
This is confirmed by our results shown in 
Fig. 2, since 
the point $B$ lies 
on the arc $AC$. When 
the FRW model is generalized to the 
anisotropic Bianchi type V universe, we find that
only a relatively small region of the 
ellipse in Fig. 2 intersects the pre--big bang 
regime. Solutions exist in a region near to the FRW model $B$, 
where the anisotropy is small, but far from this point, 
superinflation does not proceed. 
In this sense, therefore, Fig. 2  
provides in principle a quantitative 
measure of the 
likelihood of pre--big bang inflation
for generic values of $(m,p)$ in the type V
cosmology,
given by the relative size of the arc
AC to the allowed sections of the corresponding ellipse.

Finally, the ellipse (\ref{ellipse}) 
is also shown in  Figs. 1 and 2
when $\omega = \{ 0 , -4/3 , -3/2 \}$. The 
first two values represent 
the upper and lower bounds on 
$\omega$  consistent with pre--big bang inflation
in the isotropic models \cite{lidsey}. In the anisotropic models 
considered here, the $\omega =0$ curve just intersects 
the bold--faced regions and this value again 
represents the upper limit on $\omega$ for the
presence of superinflation.
However, the lower bound 
is decreased to $\omega > -3/2$. (We 
do not consider $\omega  <-3/2$ 
because this leads to a violation of the 
weak energy condition). It is interesting 
that the range of $\omega$ is broadened in these 
more general models and includes a simple class of 
higher--dimensional, Kaluza--Klein cosmologies \cite{freund}. 

In the following Section
we derive bounds for successful pre--big bang inflation
in the string--inspired model on the 
gravitational coupling
and spatial curvature. 

\section{Conditions for successful inflation}

\setcounter{equation}{0}
 
\def\theequation{\thesection.\arabic{equation}}

In the Bianchi type ${\rm VI}_h$ models 
considered in the previous Section, the Brans--Dicke field 
begins to move move away from its asymptotic value  
when $\eta ={\cal{O}} (-1)$. Its kinetic energy then  
increases significantly and this 
epoch denotes the onset of inflation. Since 
$p>0$, the 
effective gravitational coupling, $G \propto \Phi^{-1}$, 
diverges in the limit $\eta \rightarrow 0^-$. 
The question that now arises is whether 
sufficient inflation occurs before the regime of 
strong coupling and high space--time 
curvature is attained. Turner and Weinberg 
have derived bounds on the Brans--Dicke field for 
sufficient inflation in the open FRW model and have shown that the 
universe must be in the very small coupling and curvature 
regimes at the onset of 
the pre--big bang inflationary epoch \cite{tw}. Similar conclusions 
have been drawn in the inhomogeneous case \cite{ven1}. 
We generalize these bounds to anisotropic, negatively curved  universes. 

The discussion following Fig. 2 implies that the 
initial anisotropy should not be too 
pronounced and, consequently, we assume that 
$a_{1}  = {\cal{O}}( a_{2}) =
{\cal{O}} (a_{3})$ when $\eta ={\cal{O}} (-1)$. We
further assume that spatial curvature rapidly becomes 
negligible once inflation begins and that the Bianchi type 
I solution (\ref{I}) applies for $\eta > -1$. 
The amount of inflation 
that occurs in each direction 
is given  by the ratio
\be
\label{Z}
Z_i \equiv \frac{H_{if}a_{if}}{H_{ib}a_{ib}}  ,
\ee
where subscripts `$b$' and `$f$' denote values 
at the onset and end of inflation, respectively. Substitution 
of Eq. (\ref{I}) into Eq. (\ref{Z}) implies that
\be
\label{simpleZ}
Z_i \approx \left( \frac{\eta_b}{\eta_f} \right)^{1+p_1-p_i}
\approx \left( \frac{\Phi_b}{\Phi_f} \right)^{(1+p_1-p_i)/p}
\approx \left( \frac{H_{if}}{H_{ib}} \right)^{(1+p_1-p_i)/(1+p_1)}  ,
\ee
where $\Phi_b \approx Q^2$. 
In general, the inflationary solution derived from the tree--level  
string effective action breaks down at $t \approx t_f$ when 
either the coupling exceeds the 
critical value $\Phi^{-1} (t_f) \approx l_{\rm st}^{2}$ or
the curvature becomes comparable to the string scale, 
$H^{2}_i (t_f) \approx l_{\rm st}^{-2}$.
Substituting these conditions into Eq. (\ref{simpleZ}) 
then implies that 
\be
\label{minZ}
Z_i < {\rm Min} \left\{ \left( Q^2l_{\rm st}^2 
\right)^{(1+p_1-p_i)/p}, \left( H_{ib}l_{\rm st} 
\right)^{(p_i-1-p_1)/(1+p_1)} \right\} .
\ee

Since $\eta_b = {\cal{O}}(-1)$, it follows that 
$\Phi (-\eta_b)  \approx -\Phi' (-\eta_b)$. Moreover, 
approximating the solution to that of the type I cosmology 
implies that 
$H_i \approx -p_i/[A_1 (-\eta)^{1+p_1}]$, 
and, therefore, that $H_{ib} \approx 1/A_1$. 
The definition (\ref{sep3}) then implies that 
$H_{ib} \approx (Q^2/B)^{1/2}$. Substituting these 
relationships into Eq. (\ref{minZ}) and 
combining the two inequalities leads to an 
upper limit on the amount of pre--big bang inflation 
in each of the three spatial directions: 
\be
\label{upperlimit}
Z_i \le B^{r_i} \approx \left(- \dot{\Phi}_b a^3_b \right)^{r_i}, 
\qquad r_i \equiv 
\frac{1+p_1 -p_i}{2(1+p_1) +p}   ,
\ee
where a dot denotes differentiation with 
respect to cosmic time, $t$. 

This expression generalizes the result of Turner and Weinberg \cite{tw}
to the class of negatively curved, anisotropic cosmological models 
that exhibit similar qualitative behaviour to that 
of the type ${\rm VI}_h$ model.
The effect of the anisotropy 
on the maximum amount of inflation is parametrized 
completely in terms of 
the exponents $r_i$. The constraints (\ref{const3}) 
ensure that these quantities are positive definite and 
they take the values $r_i = 1/3$ in the isotropic FRW limit. 
Successful inflation 
requires that all directions expand by factors of at least 
$Z_i\ge e^{60}$ \cite{inflation} and this 
implies that $B \ge {\rm exp} (60/r_i)$. 
Consequently, the spatial direction with 
the {\em smallest} exponent determines the strongest 
limit on the value of $B=-\dot{\Phi}a^3$. The constraint 
is weakest when $r_i$ are simultaneously maximized. 
Since $\sum_i r_i =1$ and $r_i >0$, 
their maximal values correspond to 
the isotropic limit, $r_i =1/3$. We conclude, 
therefore, that a 
necessary, but not sufficient, 
condition for 
successful inflation is $-\dot{\Phi}a^3 \ge e^{180}$. 
This becomes a sufficient 
condition only in the isotropic limit. This implies that the constraints 
in anisotropic models are stronger than in the 
isotropic case.

The inflation factors in the type V model are bounded such that 
\be
\label{Vconstraint}
Z_1 \le  B^{1/3} , \qquad
Z_2 \le  B^{(2-p-m)/6}, \qquad
Z_3 \le  B^{(2+p+m)/6}  .
\ee
Either one of the exponents $r_2$ or $r_3$ is minimized when 
$|p+m|$ is maximized 
and this occurs
when  $|p+m| \rightarrow (\sqrt{5} -1)/2$.
Substituting this limiting value into the 
constraint (\ref{Vconstraint}) implies that 
$B \ge e^{260}$ is a sufficient condition 
for successful type V pre--big bang inflation. 
For the Bianchi type III model, the 
constraints on the inflation factors are 
\be
Z_1=Z_2 \le  B^{1/(4+m+p)}, \qquad
Z_3 \le B^{(2+m+p)/(4+m+p)}.
\ee
In this model, $p+m <0$, and the exponents $r_1=r_2$ are 
maximized (minimized) when $|p+m|$ 
is maximized (minimized). It follows from Eqs. (\ref{ellipse}) 
and (\ref{const3}) that the minimum values the exponents 
may take are $r_1=r_2 = 0.25$ and $r_3 = 0.23$. Thus, the
corresponding sufficient bound for successful inflation 
in this model is $B \ge e^{240}$. 

Finally, instead of 
imposing initial conditions in the limit $t \rightarrow -\infty$, 
one may consider a 
scenario where quantum stringy effects result in the 
spontaneous creation of a sufficiently homogeneous 
universe at a time $\eta = \eta_0$ \cite{tw}. In this case, 
inflation will not begin immediately
if $|\eta_0| > | \eta_b|$ and Eq. (\ref{upperlimit}) 
then determines the total amount of inflation 
that is possible.

If, on the other hand, $|\eta _0|  < |\eta_b|$, 
inflation begins immediately, but 
Eq. (\ref{upperlimit}) overestimates 
the amount of inflation because the expansion 
during the epoch $\eta_b < \eta < \eta_0$ does not 
occur. Thus, 
the actual amount of inflation is given 
by 
\be
\label{actual}
Z_{i, {\rm actual}} = \frac{H_{ib}a_{ib}}{H_{i0}a_{i0}} Z_i 
\approx \left( \frac{\eta_b}{\eta_0} \right)^{p_i-1-p_1} Z_i   .
\ee
By employing the relations  
$a^2 \Phi \propto (-\eta )^{(2+2p_1+p)/3}$ 
and $B \approx a^2_b \Phi_b$, together with Eqs. 
(\ref{sep3}) and 
(\ref{upperlimit}), we find that 
\be
\label{upperactual}
Z_{i, {\rm actual}} \le \left( 
\frac{\Phi_0^3}{\dot{\Phi}^2_0} \right)^{r_i}  .
\ee
Thus, the coupling must be sufficiently small 
at $\eta =\eta_0$ if enough inflation is to occur. 

\section{Past asymptotic state of the universe}

\setcounter{equation}{0}
 
\def\theequation{\thesection.\arabic{equation}}

It is important to consider the initial state of the universe as $t
\rightarrow -\infty$. Recently, Buonanno {\em et al.} have conjectured
that with sufficiently isotropic initial data, 
the Milne universe represents an attractor in the far past for
negatively curved cosmologies 
\cite{ven1}.  The Milne universe represents the wedge of
Minkowski space--time corresponding to the future (past) light--cone,
$|t| > |{\rm {\bf x}}|$, of the origin. In this sense, it
corresponds to the string perturbative vacuum and its spatial
three--sections, $t={\rm constant}$, are constant curvature hyperbolic
surfaces \cite{bdbook}.  The exact, negatively curved FRW solutions
with radiation and non--trivial axion, dilaton and moduli fields
approach the Milne solution in the infinite past \cite{tw,ed}.
 
In this Section we determine the initial state of the class 
of type ${\rm VI}_h$ cosmologies (\ref{sol1})--(\ref{sol4}) within 
the context of the pre--big bang scenario without  
restricting the level of anisotropy. 
As $\eta  \rightarrow -\infty$, 
the dilaton tends to a constant value
and the metric (\ref{6hmetric}) reduces to the vacuum 
solution of Lifshitz and Khalatnikov \cite{lk}: 
\bea
\label{asymmet}
ds^2 =-dt^2 +(1+k^2)^2(-t)^2dx^2 +e^{-2(1+k) x}
(-t)^{2(1+k)/(1+k^2)} dy^2 \nonumber \\
+e^{2(k-1)x} (-t)^{2(1-k)/(1+k^2)} dz^2  ,
\eea
where the variables $\{ y ,z \}$ have been rescaled without 
loss of generality. Defining the coordinate pair
\be
\label{uv}
u \equiv (-t) e^{-(1+k^2)x}, \qquad 
v \equiv (-t)e^{(1+k^2)x}
\ee
transforms the metric (\ref{asymmet}) to the null form
\be
\label{plane}
 ds^2 = -dudv+ u^{2(1+k)/(1+k^2)} dy^2 + u^{2(1-k)/(1+k^2)} dz^2
\ee
and a further change of variables: 
\bea
\label{hvar}
v \equiv  \bar{v} -\frac{1}{u(1+k^2)} \left( 
\bar{y}^2 +\bar{z}^2 +2k \bar{y} \bar{z} \right) \nonumber \\
y \equiv \frac{1}{\sqrt{2}} u^{-(1+k)/(1+k^2)} (\bar{y} 
+\bar{z} ) \nonumber \\
z \equiv \frac{1}{\sqrt{2}} u^{-(1-k)/(1+k^2)} (\bar{y}-\bar{z })
\eea
implies that 
\be
\label{planewave}
ds^2 =-dud\bar{v} +F(u, \bar{y}, \bar{z})
du^2 +d\bar{y}^2 +d\bar{z}^2  ,
\ee
where
\be
\label{HU}
F(u,\bar{y}, \bar{z}) \equiv H(u)\bar{y}\bar{z}, 
\qquad H(u) \equiv \frac{2k(1-k^2)}{(1+k^2)^2} u^{-2}   .
\ee

In general, Eq. (\ref{planewave}) is the metric for a 
gravitational plane wave 
with an amplitude uniquely 
determined by the function  $H(u)$ \cite{siklos78,lk,b}.
(For a review of the properties of plane wave metrics see, e.g., 
Ref. \cite{kshm}). These metrics are Ricci--flat and exhibit 
a covariantly constant null Killing vector field,
$l_{\mu} \equiv \partial_{\mu} u$, where  
$l^{\mu} 
\partial_{\mu} \bar{v} =1$, and the 
Riemann curvature is given by 
$R_{\mu\nu\rho\sigma} = 2l_{[ \mu} \partial_{\nu ]} \partial_{[\rho }
F l_{\sigma ]}$. It follows that the only non--zero 
component is $R_{u\bar{y} u \bar{z}} = -\frac{1}{2} H(u)$
and for $k \ne \{ 0,1 \}$,
this tends to zero as $|u| \rightarrow +\infty$.  

It follows that in the limit of $\eta  \rightarrow -\infty$, where
the metric (\ref{6hmetric}) reduces to (\ref{asymmet}),
the Riemann curvature is 
{\em identically} zero for the Bianchi types III and V. 
Thus, the initial state of 
these models is isomorphic to 
the string perturbative vacuum and 
substituting the new variables
\bea
\label{newt}
-t &\equiv & \left[ T^2 -X^2 -Y^2 -(1-k) Z^2 \right]^{1/2} \\
\label{newx}
x &\equiv & \frac{1}{2(1+k)} \ln \left[ 
\frac{T^2-X^2-Y^2-(1-k) Z^2}{(T-X)^2} \right] \\
\label{newy}
y &\equiv & \frac{Y}{T-X} \\
\label{newz}
z &\equiv & \frac{Z}{(T-X)^{1-k}}
\eea
into Eq. (\ref{asymmet}) transforms the metric into 
the standard Minkowski
form, $d\eta^2 =-dT^2+dX_idX^i$, when $k= \{ 0,1 \}$. 

It is important to note
that the entire Minkowski space--time 
is not covered by the set of 
coordinates (\ref{newt})--(\ref{newz}). The type V 
model is isomorphic to the wedge representing the $(3+1)$--dimensional 
Milne universe. In the type III model, there is no restriction 
from the $Z$ coordinate on the timelike variable, $T$, 
and this background is formally 
equivalent to the product of the $(2+1)$--dimensional Milne model 
with a line. The asymptotic behaviour of these exact type III and V 
solutions are therefore consistent with 
the conjecture of Buonanno {\em et al.} \cite{ven1}.  

\section{Discussion and Conclusion}

\setcounter{equation}{0}
 
\def\theequation{\thesection.\arabic{equation}}

In this paper we have considered the pre--big bang 
scenario within the context of the 
spatially homogeneous, diagonal Bianchi type ${\rm VI}_{h}$ 
universes, including the types III, V and negatively curved 
FRW models as particular cases. 
We find that the combined 
effects of anisotropy and spatial curvature can 
prevent pre--big bang inflation, 
in contrast to the negatively curved 
 FRW cosmology. In the region of parameter space where inflation 
does occur, the gravitational coupling and spatial curvature must satisfy 
appropriate bounds for successful inflation. In general,  
these bounds are stronger in the anisotropic models than the FRW models. 
In particular, the constraint $\ln (-\dot{\Phi} a^3 )> 180$ 
is a sufficient condition for successful inflation in the 
FRW universe, but this is strengthened to 
 $\ln (-\dot{\Phi} a^3 )> 260$ in 
the type V model. 

The past asymptotic state of the models was established.  
In general, the asymptotic
behaviour of the scale factors (\ref{sol1})--(\ref{sol3}) would seem
to indicate that the universe must have been highly
anisotropic and infinitely 
large in the far past and it could be argued that such a state is 
unnatural \cite{tw,klb}. However, 
the type III and V models are isomorphic 
to the wedge of Minkowski space--time (string 
perturbative vacuum) corresponding to 
the Milne universe with an asymptotically constant 
dilaton field. This is in agreement with 
the postulates of the pre--big bang cosmology. 

On the other hand, we have found that 
the asymptotic limit of the generic diagonal type ${\rm VI}_{h}$ 
cosmology with a dilaton 
corresponds to a homogeneous gravitational plane wave.
This supports the conjecture that 
Bianchi type IV, ${\rm VI}_h$ and ${\rm VII}_h$ 
models containing a 
perfect fluid with pressure $p$ and 
energy density $\rho$ related by 
$p=(\gamma -1)\rho$ $(2/3 < \gamma 
\le 2)$ are asymptotic to a plane wave model or the Collins 
${\rm VI}_h$ model \cite{collins,hw}. 

There are a number of reasons why plane wave backgrounds  
represent a generic initial state for the universe in 
the pre--big bang scenario. Firstly, in the space of initial data, the
most general vacuum Bianchi type B models are the 
types ${\rm VI}_h$ 
and ${\rm VII}_h$.
All but a set of measure zero 
of the known (asymptotically self--similar) solutions belonging to
these types correspond to gravitational plane waves
\cite{siklos78,hw,lukash,Barrows}. These solutions   
have been classified in a unified manner by Siklos \cite{siklos78} 
and recently surveyed in Ref. \cite{hsuw}. 

Secondly, gravitational plane waves are manifestly Ricci--flat and 
  exhibit the important property that their 
curvature invariants vanish identically. 
Consequently, they are {\em exact} 
solutions to the classical 
string equations of motion to all orders in the 
inverse string tension \cite{guv}. 
Moreover, both the ${\rm VI}_{h}$ and ${\rm VII}_{h}$ vacuum solutions 
exhibit unbroken space--time supersymmetries with 
constant Killing spinors \cite{guv,ortin}. This is important because 
supersymmetric solutions have 
a nonrenormalization property and therefore 
play a central role in quantum theories of gravity \cite{super}. It 
has been suggested that supersymmetric plane waves may
provide a suitable basis for an expansion of the path integral in 
quantum string gravity and such an approach may yield further   
insight into the high curvature, strong--coupling regime of the 
pre--big bang scenario. 
 
Finally, higher dimensions 
are important in any realistic string cosmology. Recent advances  
in string theory indicate that the five separate 
theories have a common 
origin in an eleven--dimensional `M--theory', where 
the radius of the eleventh dimension is related to the dilaton, 
$R_{11} \approx l_{\rm st} e^{\phi/3}$  
\cite{duality,M}. Within the context of M--theory, 
the thickness  
of the eleventh dimension should be vanishingly small
if the initial state of the universe 
is the ten--dimensional string perturbative 
vacuum with vanishing coupling $(\phi 
\rightarrow -\infty)$. Since  
the low--energy effective 
action of M--theory is eleven--dimensional 
supergravity with a vacuum limit given by Einstein 
gravity \cite{cj}, a 
possible initial state that avoids such a difficulty 
is an eleven--dimensional gravitational plane wave  
manifold that is homeomorphic to 
${\cal R}^{11}$. 

It would be interesting to consider the implications
of such an initial state further. In particular, 
the embedding of four--dimensional cosmological solutions in such an  
eleven-dimensional background could be established by 
employing the known theorems of differential 
geometry. For example, there 
exists a theorem that states that any $n$--dimensional 
Riemannian manifold 
may be locally and isometrically 
embedded in a {\em Ricci--flat}, Riemannian 
manifold of arbitrary dimension, $N \ge n+1$ \cite{campbell}. 
A number of plane wave solutions were recently embedded 
in five--dimensional 
spaces by employing such a theorem \cite{campbell1}. 

The bounds on the coupling and curvature derived in this paper 
correspond to the string model, $\omega =-1$, and it would 
be interesting to generalize the results to other values 
of $\omega$. It would also be of interest 
to extend the analysis to include other fields that 
arise in the low--energy string effective action. 
Homogeneous solutions with a non--trivial 
Neveu--Schwarz/Neveu--Schwarz (NS--NS) form field 
have been found for a wide variety of Bianchi and Kantowski--Sachs 
models \cite{ed,batakis,barrow}. These solutions may serve as a basis 
for investigating whether the bounds for successful inflation 
in highly anisotropic models are altered 
when such a form field is present. Their asymptotic limits 
could also be investigated to establish whether class B 
cosmologies containing both a dilaton and NS--NS form field
asymptotically approach a plane wave model.

In conclusion, therefore, spatial curvature and anisotropy 
lead to a number of important physical effects 
in the pre--big bang scenario. 

\vspace{.3in}
\centerline{{\bf Acknowledgments}}
\vspace{.3in}
DC and JEL are supported by the Particle Physics and 
Astronomy Research Council (PPARC), UK. JEL 
thanks the Astronomy Unit, Queen Mary and Westfield, and the 
Theory Division, CERN, for hospitality. We
thank J. D. Barrow, N. P. Dabrowski, 
J. Maharana, S. T. C. Siklos, M. S. Turner and G. Veneziano 
for helpful discussions and communications.

\end{document}